\documentclass[twocolumn]{aastex62}
\usepackage{amsmath}

\usepackage{float}
\graphicspath{{./}{figures/}}

\shorttitle{Observable Signatures of the Ejection Speed of Interstellar Objects}
\shortauthors{Siraj \& Loeb}
\begin{document}
\title{Observable Signatures of the Ejection Speed of Interstellar Objects from their Birth Systems}

\email{amir.siraj@cfa.harvard.edu, aloeb@cfa.harvard.edu}

\author{Amir Siraj}
\affil{Department of Astronomy, Harvard University, 60 Garden Street, Cambridge, MA 02138, USA}

\author{Abraham Loeb}
\affiliation{Department of Astronomy, Harvard University, 60 Garden Street, Cambridge, MA 02138, USA}

%\keywords{asteroids: individual (A/2017 U1)}

%% depreciated in this version as it is no longer . AASTeX 


\begin{abstract}

`Oumuamua and Borisov were the first two interstellar objects confirmed in the Solar system. The upcoming commencement of the Vera C. Rubin Observatory's Legacy Survey of Space of Time (LSST) will enhance greatly the discovery rate of interstellar objects. This raises the question, what can be learned from large-number statistics of interstellar objects? Here, we show that discovery statistics provided by LSST will allow low- and high-ejection-speed populations to be distinguished using the velocity dispersion and angular anisotropy of interstellar objects. These findings can be combined with physical characterizations to yield a better understanding of planetary system origin and nature.

\end{abstract}



\section{Introduction}

`Oumuamua was the first interstellar object discovered in the Solar system \citep{2017Natur.552..378M, 2018Natur.559..223M}. Its discovery was followed by interstellar comet Borisov \citep{2020NatAs...4...53G}, and preceded by interstellar meteor CNEOS-2014-01-08 \citep{2019arXiv190407224S}. The Vera C. Rubin Observatory's Legacy Survey of Space and Time (LSST)\footnote{https://www.lsst.org/} will greatly increase the cadence of interstellar object discovery to a rate of several per year for `Oumuamua-size objects \citep{2019ApJ...884L..22R}. This raises the question, what can be learned from the large-number statistics of interstellar objects provided by LSST?

If interstellar objects originate from stars, their velocity distribution combines the stellar and the ejection speed velocity distributions. Since ejection speeds are comparable to pre-ejection orbital speeds, they reflect the region from which interstellar objects are likely produced in planetary systems. This information is essential for understanding how interstellar objects are produced. 

Because of the motion of the Sun relative to local stars, there should be an anisotropy in the distribution of interstellar objects. Here, we show that the large-number statistics of interstellar objects provided by LSST will allow for the ejection speed of interstellar objects from their parent stars to be determined, yielding important insights into planetary system evolution. Since the objects under consideration are lightweight, their gravitational deflection by stars along their journey results in a negligible change to their speed. Similarly, friction on the interstellar medium can be ignored \citep{2018ApJ...868L...1B, 2018ApJ...860...42H, 2020ApJ...899L..23H}.

Our discussion is structured as follows. In Section \ref{sec:k}, we explore the predicted kinematics of interstellar objects and describe our numerical Monte Carlo model. In Section \ref{sec:r}, we report our results. Finally, in Section \ref{sec:d} we explore key predictions and implications of our tests.

%\newpage 

\section{Kinematics and Numerical Model}
\label{sec:k}

If interstellar objects originate from planetary systems around stars, they should reflect the velocity distribution of stars. The Sun's velocity vector relative to local standard of rest (LSR) is, $(v_U^{\odot}, v_V^{\odot}, v_W^{\odot}) = (10 \pm 1, 11 \pm 2, 7 \pm 0.5) \; \mathrm{km \; s^{-1}}$ \citep{2010MNRAS.403.1829S, 2015ApJ...809..145T, 2016ARA&A..54..529B}. We note that adopting alternative definitions of the LSR, such as only averaging over dwarf stars, may alter the results described here by $< 20 \%$ \citep{2017RNAAS...1...21M}. The velocity dispersion of local stars around LSR is, $(\sigma_U, \sigma_V, \sigma_W) = (33 \pm 4, 38 \pm 4, 23 \pm 2) \; \mathrm{km \; s^{-1}}$ \citep{2018MNRAS.474..854A}. The errors quoted above represent 1-sigma uncertainties. Since planetary disks are randomly oriented, the directions of interstellar object ejection are taken to be random. 

We created a numerical Monte Carlo model that draws three-dimensional velocity components relative to the LSR from the aforementioned distributions. The model draws spherical angles $\phi$ and $\theta$ from uniform distributions to construct a randomly oriented ejection velocity vector, which is added to the velocity vector of the star, and then derives the angle of displacement between the star's velocity vector and the Solar antapex velocity. The uncertainties on the final distributions, indicated as thick bands in Figure \ref{fig:diffsv} and Figure \ref{fig:diff}, are derived through a similar Monte Carlo procedure, by drawing from the Gaussian error distribution on each velocity and dispersion component and propagating the effects accordingly.

\section{Results}
\label{sec:r}

We applied our model to stellar ejection speeds of $0$ and $50 \; \mathrm{km \; s^{-1}}$, which bracket the plausible range of values for planetary systems. Objects in the outer reaches of planetary systems would be easily freed by passing stars and the Galactic tide due to their low gravitational potential relative to their parent stars, resulting in negligible ejection speeds. On the other hand, gravitational kicks from interactions with planets in or near the habitable zones of other stars could eject a large number of planetesimals at $50 \; \mathrm{km \; s^{-1}}$, representing the orbital speed in the habitable zone of the most abundant class of stars, M dwarfs. For the population of white dwarfs (as suggested for an origin of `Oumuamua by \citealt{2018ApJ...861...35R}), the ejection speed in the habitable zone could be an order of magnitude bigger and should be easy to recognize in the tail of the velocity distribution.

\begin{figure}
  \centering
  \includegraphics[width=1\linewidth]{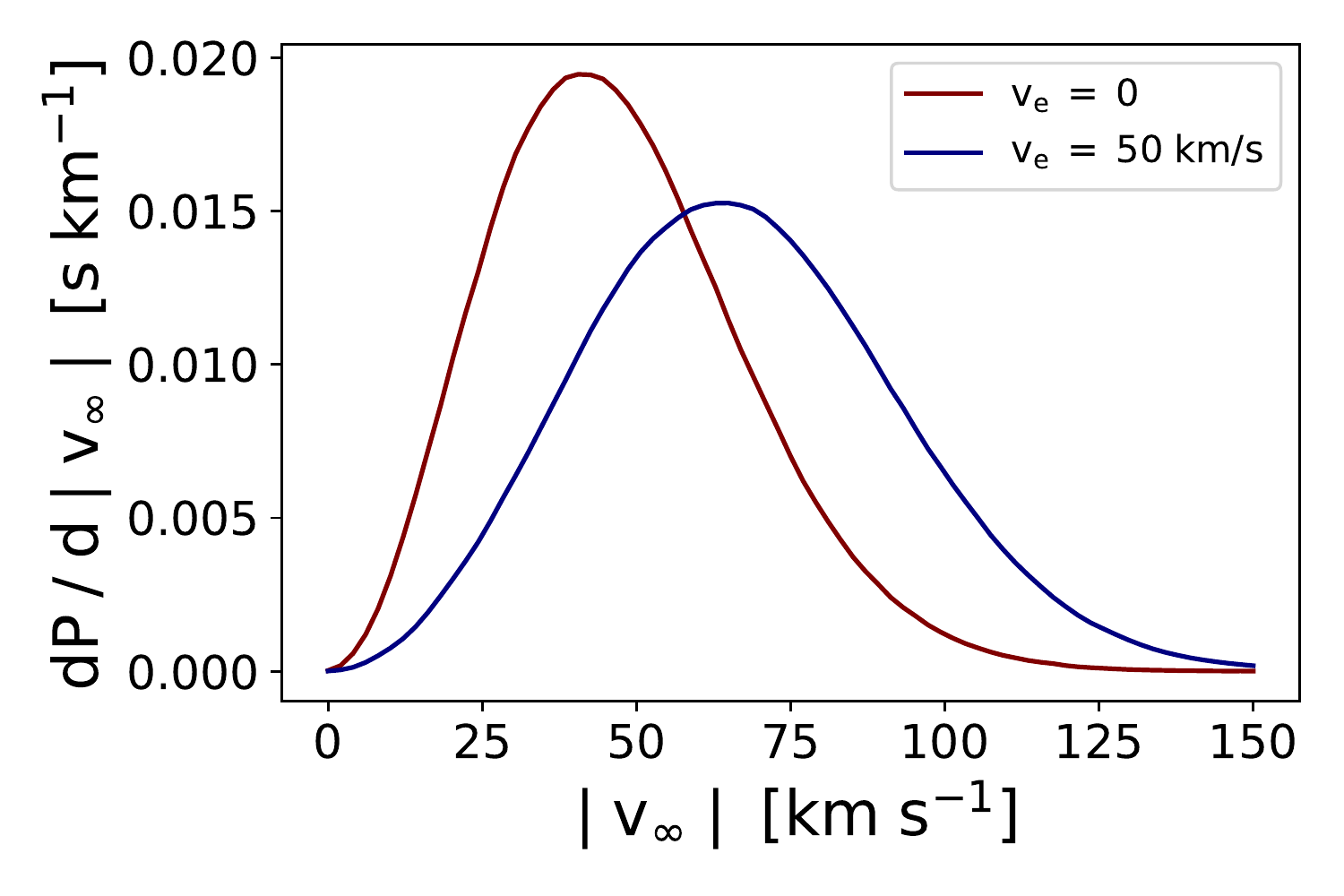}
    \caption{Normalized probability distributions for the absolute magnitude of the speed of interstellar objects relative to the Sun $v_{\infty}$, given the stellar ejection speed cases of $0$ and $50 \; \mathrm{km \; s^{-1}}$.
}
    \label{fig:v1}
\end{figure}

\begin{figure}
  \centering
  \includegraphics[width=1\linewidth]{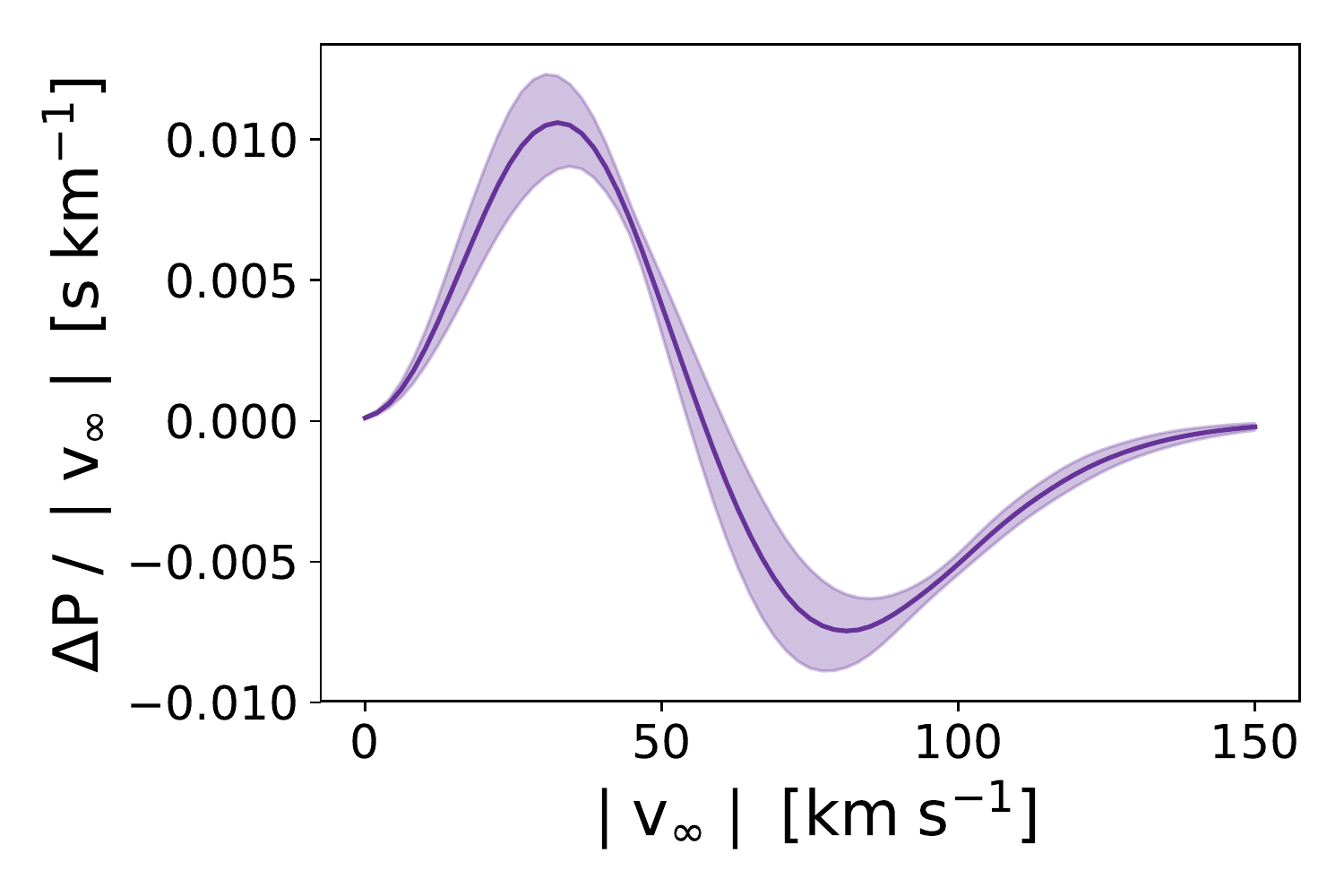}
    \caption{Difference between the probability distributions shown in Figure \ref{fig:v1}, with the shaded one-sigma error band corresponding to uncertainties in the underlying LSR velocity and dispersion components.
}
    \label{fig:diffsv}
\end{figure}

Figure \ref{fig:v1} displays the distributions for the speed of interstellar objects $v_{\infty}$ relative to the Sun given ejection speeds of $0$ and $50 \mathrm{\; km \; s^{-1}}$, as computed by the Monte Carlo model described in Section \ref{sec:k}. Figure \ref{fig:diffsv} shows the difference between the two distributions in Figure \ref{fig:v1}, with the one-sigma uncertainties displayed as the shaded band. The $v_{\infty}$ bins near $\sim 30 \mathrm{\; km \;s^{-1}}$ and $\sim 80 \mathrm{\; km \;s^{-1}}$ are best for discerning between the ejection speed distributions.

\begin{figure}
  \centering
  \includegraphics[width=1\linewidth]{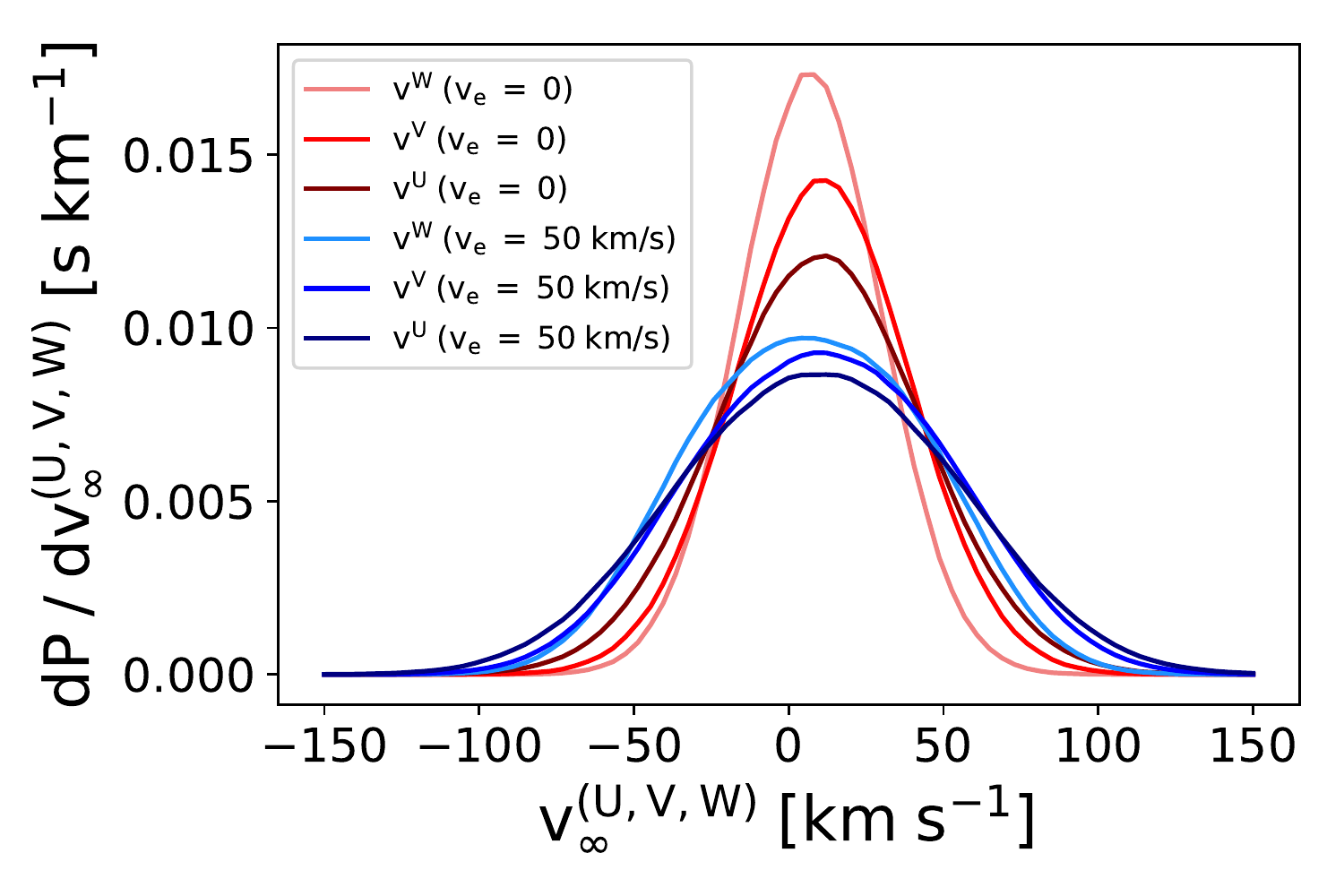}
    \caption{Normalized probability distributions for $v^{U}_{\infty}$, $v^{V}_{\infty}$, and $v^{W}_{\infty}$, given the ejection speed scenarios $0$ and $50 \; \mathrm{km \; s^{-1}}$.
}
    \label{fig:v2}
\end{figure}

Figure \ref{fig:v2} shows the distributions of $v_{\infty}$ separated by velocity components in the Galactic directions U, V, and W, which, added in quadrature, yield the distributions in Figure \ref{fig:v1}.

\begin{figure}
  \centering
  \includegraphics[width=1\linewidth]{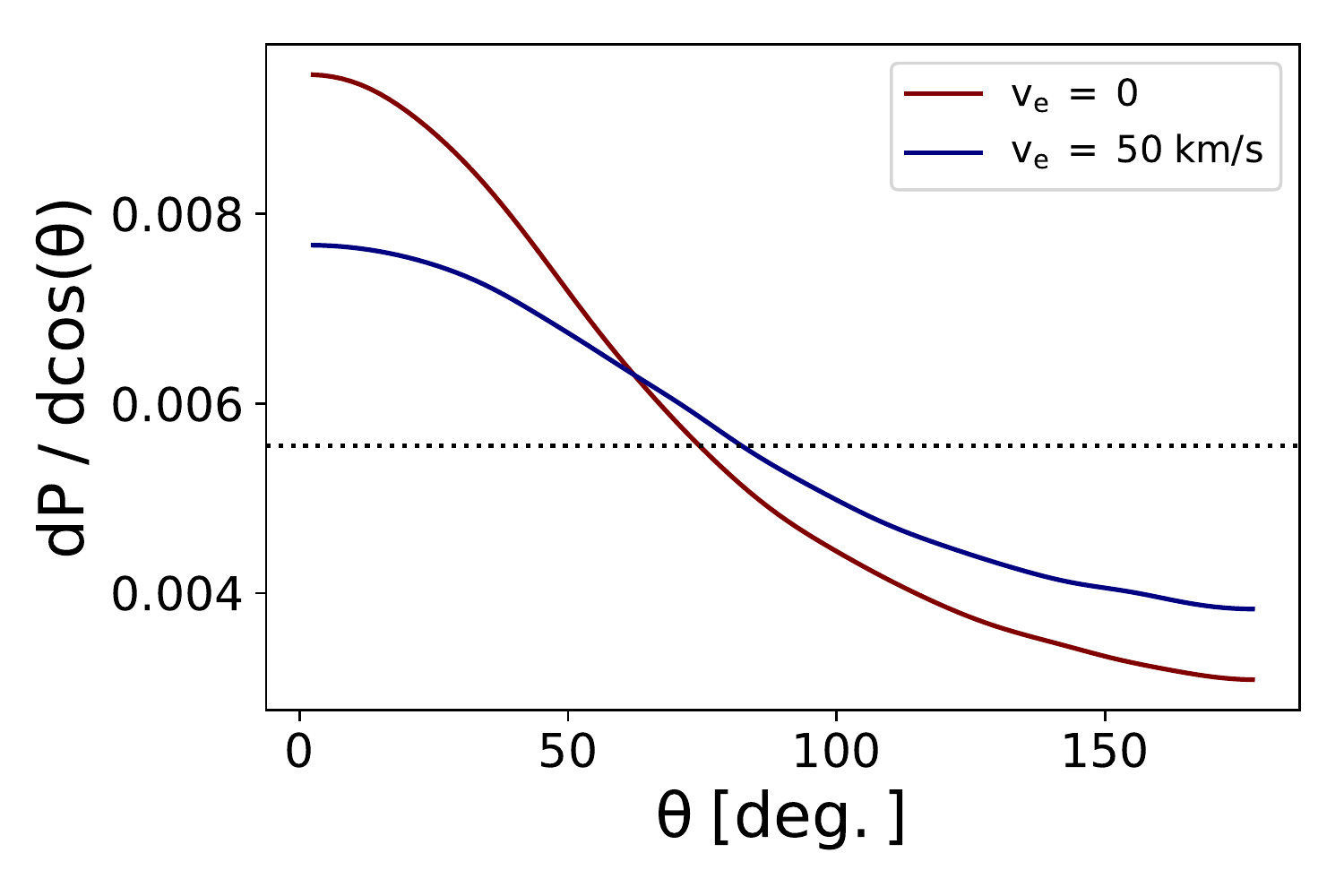}
    \caption{Normalized differential probability distributions, for ejection speeds of $0$ and $50 \; \mathrm{km \; s^{-1}}$, as a function of angular velocity vector displacement from the Solar antapex. If the Sun were stationary to the LSR, the distribution of $dP / d \cos{(\theta)}$ would be uniform, as indicated by the dotted line.
}
    \label{fig:eject}
\end{figure}

Figure \ref{fig:eject} shows the normalized probability distributions, adjusted for the solid angle associated with $\theta$ for the $0$ and $50 \mathrm{\; km \; s^{-1}}$ ejection speed cases. A systematically larger ejection speed leads to a flattened distribution of $d\mathrm{P} / d \cos{\mathrm{\theta}}$.

\begin{figure}
  \centering
  \includegraphics[width=1\linewidth]{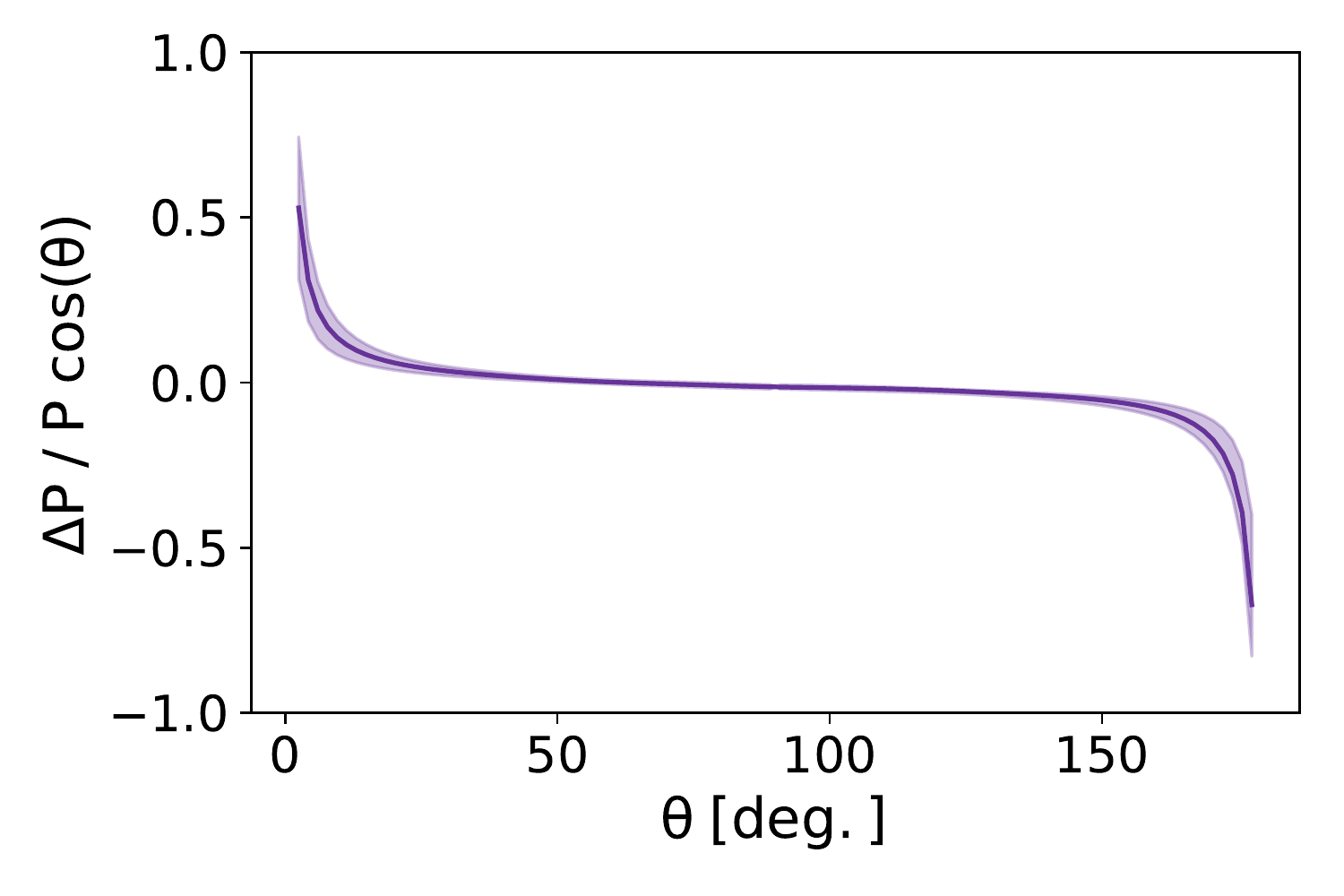}
    \caption{Relative difference between the two probability distributions shown in Figure \ref{fig:eject}, with the shaded one-sigma error band corresponding to uncertainties in the underlying LSR velocity and dispersion components.
}
    \label{fig:diff}
\end{figure}

Figure \ref{fig:diff} indicates the relative difference between the $0$ and $50 \mathrm{\; km \; s^{-1}}$ ejection speed scenario probability distributions, with one-sigma uncertainties displayed as a shaded band. The angular bins near the Solar antapex and the Solar apex are best for discerning between ejection speed distributions. Interstellar objects with velocity vectors near the Solar antapex, which oppose the motion of the Sun, are the favored population to differentiate between ejection speed distributions since their abundance is expected to be at least a factor of $\sim 2$ greater than interstellar objects with velocity vectors near the Solar apex.

\section{Discussion}
\label{sec:d}

We showed that there is expected to be an anisotropy in the velocity phase space of interstellar objects owing to the Sun's motion relative to the LSR. Furthermore, we demonstrated that a larger ejection speed of interstellar objects from their parent stars will lead to an increase in the predicted $v_{\infty}$ distribution and reduction in the predicted angular anisotropy. The difference between a low- and high-ejection-speed interstellar objects will be observable by LSST and the overall degree of angular anisotropy will be revealed best by interstellar objects with velocity vectors similar to that of the Solar antapex. With a sufficiently large number of interstellar objects, it should be possible to infer the characteristic value of $v_e$. 

Under the null hypothesis ($v_e = 0$ ), $\sim 50 \%$ of objects will have speeds of of $v_{\infty} \lesssim 40 \mathrm{\; km \; s^{-1}}$. Given the constraint $v_{\infty} \lesssim 40 \mathrm{\; km \; s^{-1}}$, there is a $\sim 60\%$ difference in the distributions of $dP / d \lvert v_{\infty} \rvert$ between the two ejection scenarios, implying through Poisson statistics that a total of $\sim 12$ interstellar objects will allow for the null hypothesis to be confirmed or rejected relative to the alternative hypothesis ($v_e = 50 \; \mathrm{km \; s^{-1}}$) at a two-sigma confidence level. Similarly, $\sim 50 \%$ of objects will have angular displacements of $\theta \lesssim 60^{\circ}$, and given this constraint there is a $\sim 15\%$ difference in the distributions of $dP / d \cos{(\theta)}$ between the two ejection scenarios, and a $\sim 60\%$ between the $v_e = 0$ ejection scenario and a hypothesis that predicts zero angular anisotropy. Poisson statistics indicate that $\sim 180$ interstellar objects are necessary for the null to be confirmed or rejected relative to the alternative hypothesis ($v_e = 50$) at a two-sigma confidence level, and that $\sim 12$ interstellar objects will allow for the confirmation of the existence of angular anisotropy in the distribution of interstellar objects. The discovery of tens of interstellar objects will both allow for differentiation between the $v_e = 0$ and scenarios $v_e = 50 \; \mathrm{km \; s^{-1}}$ in addition to assessing whether or not the predicted angular anisotropy exists.

Given LSST's limiting magnitude of 24.5, an `Oumuamua-size ($\sim 100 \mathrm{\; m}$) object is detectable out to a distance of $\sim 2 \mathrm{\; AU}$ \citep{2019ApJ...872L..10S}. Since the number density of `Oumuamua-size objects is $\sim 0.2 \; \mathrm{AU^{-3}}$ \citep{2018ApJ...855L..10D}, this implies that LSST may detect up to one `Oumuamua-size object per month. Interstellar objects of size $\sim 25 \mathrm {\; m}$ are detectable out to a distance of $\sim 1 \mathrm{\; AU}$, and adopting a cumulative size distribution power-law index of $-3$, a value consistent with constraints from both \cite{2019arXiv190603270S} and \cite{2020AJ....160...26B}, implies discovery by LSST at a rate of up to twice per week.

Interstellar asteroids, characterized by their rocky compositions, are expected to have high ejection speeds from their parent stars, and interstellar comets, characterized by their icy compositions, are expected to have low ejection speeds from their parent stars, since the two populations originate from the inner and outer regions, respectively, of planetary systems. The most natural expectation is that most interstellar objects originate from exo-Oort clouds since they represent the regions of exoplanetary systems with the lowest gravitational binding energies. As a result, these objects would be interstellar comets, with icy compositions, and with ejection speeds near zero. We note that interactions with turbulent source environments as well as the interstellar medium may provide additional constraints on interstellar object velocities \citep{2020ApJ...894...36L}.

Finally, we note that the complete velocity distribution of interstellar objects may contain multiple components that are distinguishable, with ejection speeds ranging from leaving outer Oort clouds around stars at low velocities ($v_e \sim 0$) up to escaping the tidal disruption orbital region near white dwarfs ($v_e \sim \mathrm{\; hundreds \; km \; s^{-1}}$). LSST will be capable of calibrating the relative mass fraction in each of these components. If extraterrestrial artifacts constitute a sub-population of interstellar objects \citep{2018ApJ...868L...1B}, they may also be distinguishable using the twin measures of velocity dispersion and angular anisotropy.
%\vspace*{0.3in} 
%\pagebreak
%\newpage
%\skip
%\begin{minipage}{0.45\textwidth}
\section*{Acknowledgements}
%\vspace{0.1in} 
We thank Manasvi Lingam for helpful comments on the manuscript. This work was supported in part by a grant from the Breakthrough Prize Foundation.\newline \newline \newline
%\end{minipage}

%\vspace*{0.3in} 



%% of citation expression problems, but it is  to clearly

%\newpage 

\bibliography{bib}{}
\bibliographystyle{aasjournal}



\end{document}